\documentclass[journal=jpclcd,manuscript=article,layout=traditional]{achemso}

\usepackage[version=3]{mhchem} 
\usepackage{textgreek}
\usepackage{amsmath}
\usepackage{amssymb}
\usepackage{subfig}
\usepackage{stmaryrd}
\usepackage{upgreek}
\usepackage{xcolor}

\author{Nicolas Gaudy}
\affiliation{CIRIMAT, Universit\'e de Toulouse, CNRS, Universit\'e Toulouse 3 - Paul Sabatier, 118 Route de Narbonne, 31062 Toulouse cedex 9, France}
\altaffiliation{R\'eseau sur le Stockage \'Electrochimique de l'\'Energie (RS2E), F\'ed\'eration de Recherche CNRS 3459, HUB de l'\'Energie, Rue Baudelocque, 80039 Amiens, France}
\author{Mathieu Salanne}
\affiliation{Sorbonne Universit\'e, CNRS, Physicochimie des \'Electrolytes et Nanosyst\`emes Interfaciaux, F-75005 Paris, France}
\altaffiliation{R\'eseau sur le Stockage \'Electrochimique de l'\'Energie (RS2E), F\'ed\'eration de Recherche CNRS 3459, HUB de l'\'Energie, Rue Baudelocque, 80039 Amiens, France}
\author{C\'eline Merlet}
\affiliation{CIRIMAT, Universit\'e de Toulouse, CNRS, Universit\'e Toulouse 3 - Paul Sabatier, 118 Route de Narbonne, 31062 Toulouse cedex 9, France}
\altaffiliation{R\'eseau sur le Stockage \'Electrochimique de l'\'Energie (RS2E), F\'ed\'eration de Recherche CNRS 3459, HUB de l'\'Energie, Rue Baudelocque, 80039 Amiens, France}
\email{celine.merlet@univ-tlse3.fr}

\title{Dynamics and energetics of ion adsorption at the interface between a pure ionic liquid and carbon electrodes}

\begin{document}

\makeatletter
\setlength\acs@tocentry@height{8.25cm}
\setlength\acs@tocentry@width{4.45cm}
\makeatother
\begin{tocentry}
\includegraphics[scale=1.0]{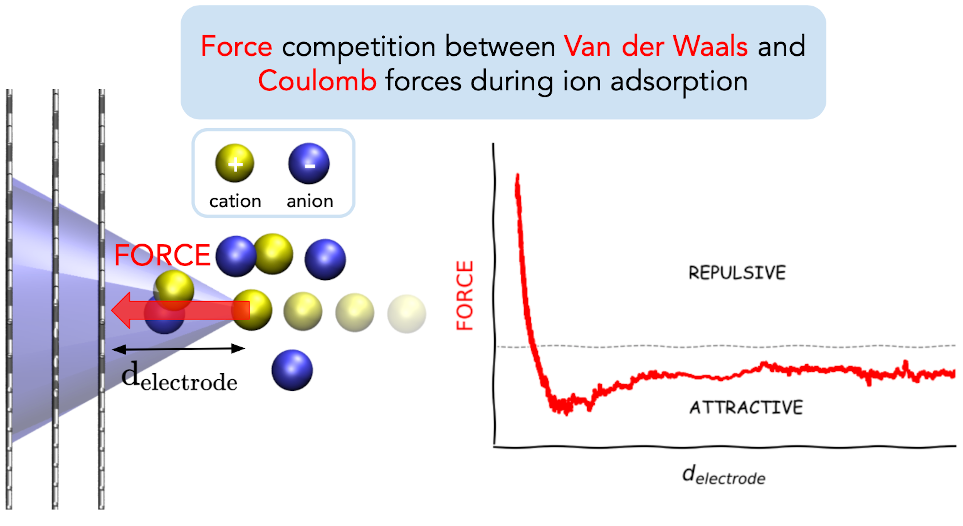}
\end{tocentry}

\begin{abstract}

Molecular dynamics simulations have been used extensively to determine equilibrium properties of the electrode-electrolyte interface in supercapacitors held at various potentials. While such studies are essential to understand and optimize the performance of such energy storage systems, investigations of the dynamics of adsorption during the charge of the supercapacitors is also necessary. Dynamical properties are especially important to get an insight into the power density of supercapacitors, one of their main assets. In this work, we propose a new method to analyze the trajectories of adsorbing ions. We focus on pure 1-ethyl-3-methylimidazolium bis(trifluoromethylsulfonyl)imide in contact with planar carbon electrodes. We characterize the evolution of the ion orientation and ion-electrode distance during adsorption and show that ions reorientate as they adsorb. We then determine the forces experienced by the adsorbing ions and demonstrate that Coulomb forces dominate at long range while van der Waals forces dominate at short range. We also show that there is an almost equal contribution from the two forces at an intermediate distance, explaining the peak of ion density close to the electrode surface. 

\end{abstract}

\section{Introduction}

Supercapacitors, also known as electrochemical capacitors, have emerged as promising energy storage devices due to their exceptional power delivery, long cycle life, and fast charge/dis\-charge capabilities\cite{miller_materials_2008}. They store energy through the formation of electrical double layers (EDLs) at the electrode-electrolyte interface, leading to the compensation of electrode charges through reversible adsorption of ions\cite{griffin_situ_2015}. Similarly to the bulk electrolyte transport properties (diffusion coefficients, electrical conductivity, viscosity), the dynamics of ion adsorption is influenced by various factors, such as electrostatic interactions, ion size and coordination. It has a direct effect on the power characteristics of capacitive systems,\cite{pean_dynamics_2014} hence it is essential to describe and understand interfacial properties dynamically at the molecular level in order to improve the performance of supercapacitors.

Classical molecular dynamics (MD) is a powerful computational tool for investigating the behaviour of supercapacitors. By performing MD simulations, it is possible to gain insights into the intricate processes occurring at the molecular level within these energy storage devices. Due to their capability to obtain trajectories for large systems ($>$~10,000~atoms) and for relatively extended duration (several nanoseconds)\cite{xu_computational_2020,park_interference_2020,jeanmairet_microscopic_2022}, these simulations provide valuable information on the mechanisms underlying charge storage, ion migration and other crucial factors that influence the efficiency and reliability of supercapacitors. In particular, MD simulations enable the exploration of various aspects of supercapacitor operation, such as ion diffusion, charge/discharge dynamics, and interfacial phenomena (EDL structure, EDL dynamic, ion orientation with respect to the electrode etc). In recent years, significant progress has been made in understanding the complex structure of ionic layers in EDLs through MD simulation studies. The effects of overscreening and ion packing on the structure of the EDL and capacitance value have been investigated\cite{fedorov_towards_2008,fedorov_double_2010,kirchner_electrical_2013,merlet_electric_2014,vatamanu_molecular_2012}. However, most of these studies have focused on the equilibrium properties of supercapacitors such as the ion density, the angular distribution of adsorbed ions, and more recently the energy aspects~\cite{de_araujo_chagas_molecular_2022,de_araujo_chagas_comparing_2023}. Although some MD simulation studies have investigated the charging dynamics of overall devices\cite{jiang_dynamics_2014,kong_molecular_2015,noh_understanding_2019,pean_dynamics_2014}, a dynamic analysis of the individual ions adsorption process has yet to be carried out. 

To address this, in this work we develop a set of methods allowing to provide a comprehensive and dynamic depiction of the ion adsorption process for a system composed of a neat IL (1-ethyl-3-methylimidazolium bis(trifluoromethylsulfonyl)imide, EMIM-TFSI) in contact with planar carbon electrodes. In particular, we focus on examining the dynamic orientation of adsorbing ions, as well as comparing the electrostatic and van der Waals ({vdW}) forces exerted on these ions throughout the charging process of the supercapacitor. By analysing the evolution of these quantities over time, we show that ions adsorb at the same time as they reorient. Once there is a vacancy in the EDL, adsorption is the result of the competition between the electrostatic and van der Waals forces.

\section{Methods}

\begin{figure}
\begin{centering}
\includegraphics[scale=0.8]{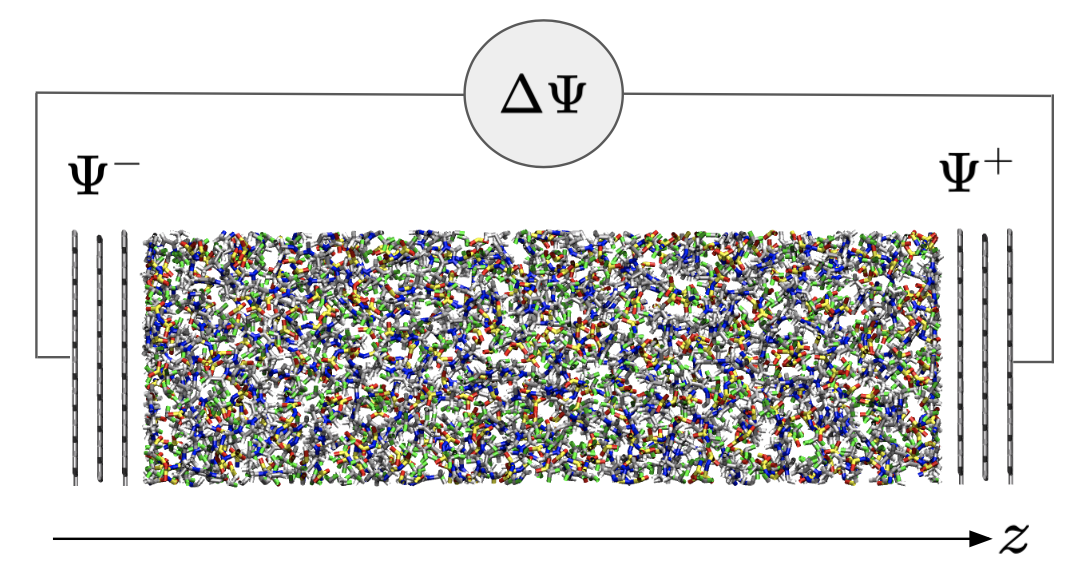}\caption{Supercapacitor composed of two graphite electrodes (in grey) in contact with a
pure ionic liquid electrolyte EMIM-TFSI. A constant potential difference is applied between the two electrodes.\label{fig:Supercapacitor-syst}}
\par\end{centering}
\end{figure}

The MetalWalls software\cite{marin-lafleche_metalwalls_2020,coretti2022a} was used to perform MD simulations of neat EMIM-TFSI in contact with two graphite electrodes, between which a constant potential difference was maintained. Figure \ref{fig:Supercapacitor-syst} provides a representative snapshot of the system. 
The electrolyte comprises 322 ion pairs, and the electrodes consist of three parallel graphene sheets that are maintained rigid and separated from each other by 3.4~\r{A}. The simulation box has dimensions of 34~$\times$~37~$\times$~112~$\textrm{{\AA}}^3$, with two-dimensional periodic boundary conditions applied in the $x$ and $y$ directions. No periodic boundary conditions are applied in the $z$ direction, perpendicular to the electrodes. 

In the constant potential method, each carbon atom $i$ of the electrode carries a Gaussian charge distribution, with a width $\eta_{i}^{-1}=0.40\,\textrm{{\AA}}$ in this work. Thus, the charge distribution of an atom in the electrode will be of amplitude $Q_{i}$ and width $\eta_{i}^{-1}$ , such that
\begin{equation}
\rho_{i}\left(\mathbf{r}\right)=Q_{i}\eta_{i}^{3}\pi^{-3/2}{\rm e}^{-\eta_{i}^{2}\left|\mathbf{r}-\mathbf{r_{i}}\right|^{2}} 
\end{equation}
where $\mathbf{r}=(x,y,z)$ is the position considered in the simulation box and $\mathbf{r_i}=(x_i,y_i,z_i)$ is the position of the $i-th$ electrode atom. The electrode atom charges fluctuate in order to satisfy the constant potential condition.~\cite{reed2007a} 

On the electrolyte side, the charges carried by the atoms are considered to be point charges. Thus, the charge distribution of an atom $i$ of charge $Q_{i}$ is given by 
\begin{equation}
\rho_{i}\left(\mathbf{r}\right)=Q_{i}\updelta\left(\mathbf{r}-\mathbf{r_{i}}\right) 
\end{equation}

To represent EMIM-TFSI, the all-atom CL\&P force field\cite{canongia_lopes_modeling_2004,canongia_lopes_molecular_2004} was employed. This force field includes electrostatic and van der Waals interactions. Atomic charges are scaled by a factor of $0.8$ to mimic polarizability effects, a technique that has been suggested in previous studies of bulk liquids\cite{bhargava_refined_2007,chaban_polarizability_2011,zhang_simple_2012,schroder2012a}. For van der Waals interactions, a Lennard-Jones potential with a cutoff of 14.8~\r{A} is implemented, while the Ewald summation method is used to calculate long-range interactions, with a real space cutoff of 17~\r{A}. The initial configuration of the system was generated using the PACKMOL software\cite{martinez_packmol_2009}. The liquid density was adjusted by allowing the electrodes to move in the $z$ direction, for a few nanoseconds, using a piston and applying a pressure of 0 bar on both sides. Once the positions of the electrodes became stable, a bulk density of $1.6\,\mathrm{g.cm^{-3}}$ was attained, which is very close to the experimental value at ambient temperature\cite{trenzado_experimental_2021}.

The system undergoes charging by applying potential differences of $\Delta\Psi$~=~1~V, 2~V and 4~V between the two electrodes. The simulations are conducted in the NVT ensemble with the temperature set to 298~K using a Nos{\'e}-Hoover thermostat chain of length 5 with a relaxation time of 500~fs. A constant applied potential is maintained between the two electrodes.~\cite{reed2007a} Initially, an equilibration phase is performed at 0~V. Following this equilibration, a production run of 63~ns is carried out, using a timestep of 2~fs. Figure~S3 shows the evolution of the electrode charge along the simulation. 

\section{Results and discussion}

In this work, we aim to investigate both structural changes and forces individually experienced by the ions during adsorption. More specifically, we focus on the adsorption of EMIM cations on the negative electrodes and TFSI anions on the positive electrodes during electrode charging with 1~V, 2~V and 4~V potential differences applied. To emphasize specific phenomena occurring during adsorption, we have developed a method for dynamically monitoring the system and statistically studying adsorption. 

\subsection{Interfacial structure and identification of the adsorbed ions}

\begin{figure}[ht]
\centering{}\includegraphics[scale=0.5]{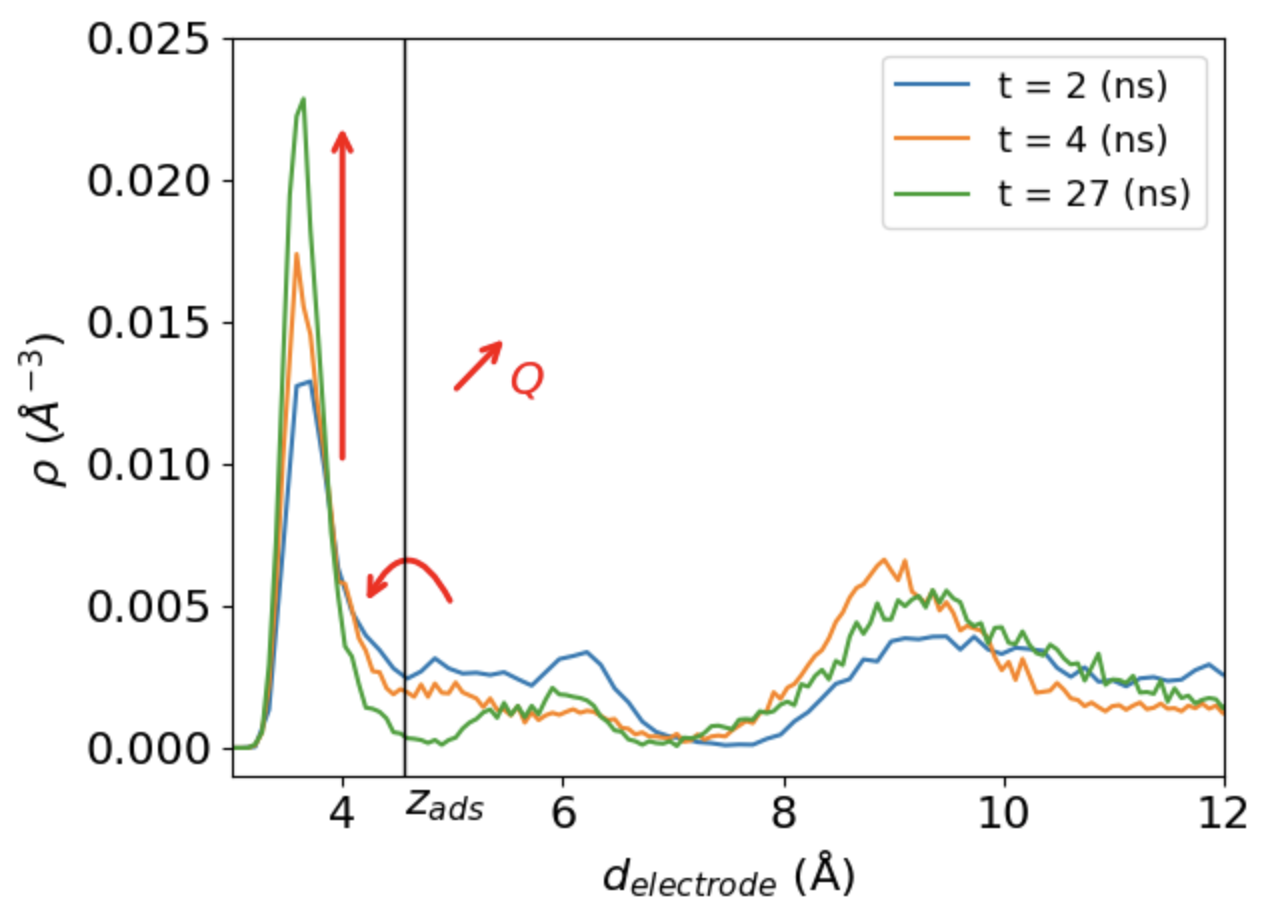}\caption{Ionic density of EMIM as a function of $d_{electrode}$ for $\Delta\Psi=1\,\mathrm{V}$. The first peak increases when the total electrode charge, $Q$, increases. The black vertical line represents the boundary between the first and second adsorption layers, noted $z_{ads}$. \label{fig:Ion-density-profile}}
\end{figure}

A common way to characterize the liquid structure at the interface with an electrode is to plot ionic densities. Figure \ref{fig:Ion-density-profile} shows the density of EMIM, $\rho$, as a function of the distance to the electrode, $d_{electrode}$, at different times during the charge of the supercapacitor. 
This distance is defined here as $z-z_{elec}$ where $z$ is the $z$-component of the position considered in the box and $z_{elec}$ is the position of the graphene sheet in contact with the electrolyte. For quantities where ionic positions are considered, $d_{electrode}$ is defined as $z_{CM}-z_{elec}$ where $z_{CM}$ is the $z$-component of the position of the center of mass of the considered ion.

There is a first intense adsorption peak for $d_{electrode}\sim$~3.6~\r{A} that increases over time. Indeed, during the charge of the supercapacitor, counterions adsorb on the electrodes and compensate their charge, which leads to an increase in the ion concentration in the first adsorbed layer. According to this ionic density profile, ions are considered to be adsorbed if $z_{CM}$ is located between the graphene sheet in contact with the electrolyte and the first minimum of the ion density, noted $z_{ads}.$ For the remainder of this article, analyzes are conducted only on ions that are not in the first adsorption layer at $t=0$ and that are there once the supercapacitor is charged, i.e. only ions which adsorb during charging.

\subsubsection{Evolution of quantities during the adsorption process}

In order to generalize the study of the adsorption process, a method for statistically studying the dynamics of adsorption was developed. During the charging of the supercapacitor, the ions that adsorb are identified and their orientation and the forces acting on them over time are computed. Subsequently, each ion is then assigned a time $t_{ads}$ such that $d_{electrode}\left(t_{ads}\right)=z_{ads}$, the time after which the ion is considered to be adsorbed. The orientations and the forces undergone by the ions are then realigned with respect to $t_{ads}$ ($t'=t-t_{ads})$. This ensures that for all the ions, the new origin $t'=0$ corresponds to the instant when the ions passes from the non-adsorbed state to the adsorbed state. In other words, for $t'<0$ the ion is considered not to be adsorbed and for $t'>0$ the ion is considered to be adsorbed. By aligning all quantities around the time of adsorption, averages of various properties can be computed and standard deviations extracted. To correlate local properties during adsorption, large vectors containing i) all the ion-electrode distances and ii) the observed quantities during adsorption are generated. It is then possible to plot any property as a function of the electrode distance and conduct subsequent analyzes. More details about this way of analyzing the trajectories done are available in SI.

\subsection{Structural description}

\begin{figure}
\begin{centering}
\includegraphics[scale=0.65]{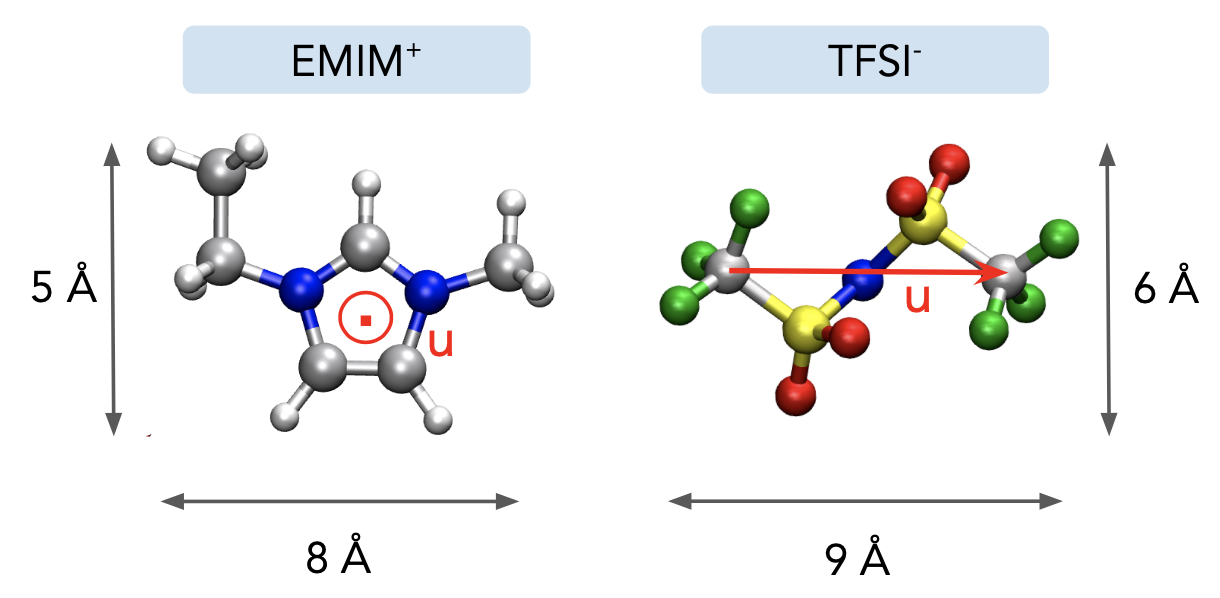}
\par\end{centering}
\caption{Scheme explaining the definition of $\mathbf{u}$ vectors for each
ion\label{fig:Diagram-orientation-vecteur}}
\end{figure}

In order to characterize the orientation of the ions with respect to the electrode on which they are adsorbed, a unit vector normal to the electrode, labelled $\mathbf{n}$, and a vector associated with each ion, noted $\mathbf{u}$, are defined. For EMIM, $\mathbf{u}$ is defined as a vector normal to the plane of the imidazolium ring. For TFSI, $\mathbf{u}$ is defined as the vector linking the two carbon atoms (Figure \ref{fig:Diagram-orientation-vecteur}). The angle $\theta=\widehat{\mathbf{n\cdot}\mathbf{u}}$ is used to determine the general orientation of the ions. For EMIM ions, if $\theta$ is between 0$^{\circ}$ and 45$^{\circ}$, the ion is considered parallel to the electrode while it is considered perpendicular to the electrode if $\theta$ is between 45$^{\circ}$ and 90$^{\circ}$. For TFSI ions, if $\theta$ is between 0$^{\circ}$ and 45$^{\circ}$, the ion is considered perpendicular to the electrode while it is considered parallel to the electrode if $\theta$ is between 45$^{\circ}$ and 90$^{\circ}$. Although stricter definitions could be used, it would not change the conclusions obtained.

\begin{figure}[h!]
\begin{centering}
\includegraphics[scale=0.5]{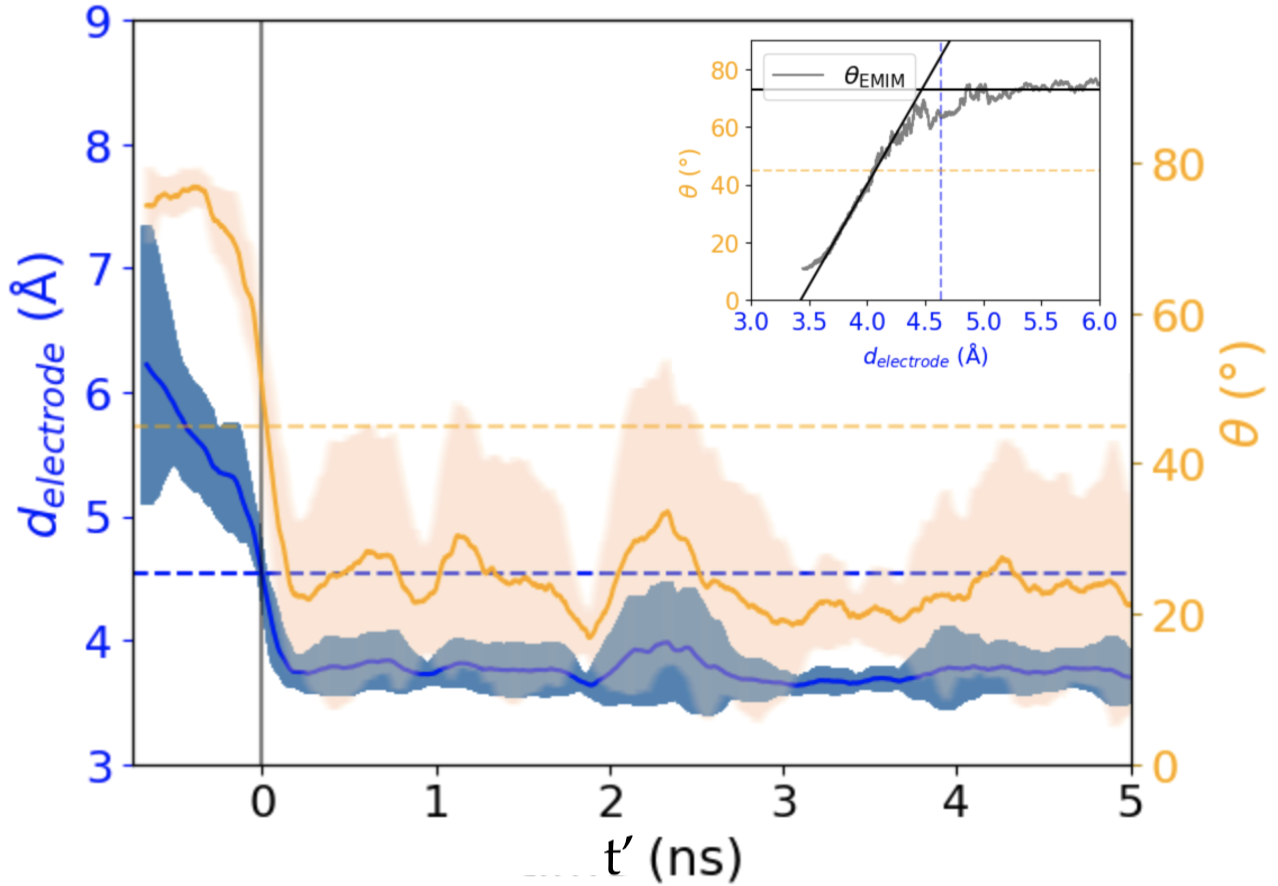}
\par\end{centering}
\caption{In blue, average distance between the electrode and the centre of mass of the EMIM ions that adsorb over time. In orange, average orientation of the EMIM ions that adsorb over time. Shaded regions indicate the range of
statistical uncertainty. The applied potential is equal to $\Delta\Psi=1\,\mathrm{V}$. The blue dashed line corresponds to $z_{ads}$, the orange dashed line corresponds to $\theta=45{^\circ}$ and the vertical black line corresponds to $t=0$ ns. Insert: average angle as function of the distance. \label{fig:ads_EMIM_dist_elec_orientation}}
\end{figure}

Figure~\ref{fig:ads_EMIM_dist_elec_orientation} shows the average distance and orientation of EMIM ions as a function of time, during adsorption, for $\Delta\Psi=1\,\mathrm{V}$. The insert shows the variation of the angle as a function of the distance from the electrode. 
It can be seen that before adsorption all the ions approach the electrode in a perpendicular orientation. Indeed, for $d_{electrode}$ between 5\r{A} and 6\r{A}, there is no variation in the angle $\theta$. When adsorption occurs, the ions reorient themselves parallel to the electrode at the same time as they enter the first adsorbed layer. Once the ions are adsorbed, the angle varies linearly with the variation in distance from the electrode. As the ions reorient, their centre of mass moves and so their distance from the electrode decreases. Figure~S4 reports the raw data for distances and angles, before averaging, showing the extent of the deviations observed for $d_{electrode}$ and $\theta$. The same behaviour was observed for all potential differences and for TFSI anions.

\subsection{Forces description}

Following the analysis of the interfacial structure of the ions, the forces felt by the ions during the adsorption process are determined. Specifically, we investigate electrostatic interactions, which are characterized by the Coulomb force, as well as van der Waals interactions, which are derived from the Lennard-Jones potential. The force derived from a potential exerted by a particle $j$ on a particle $i$ is given by 
\begin{equation}
\mathbf{F_{ij}}=-\nabla\mathcal{U}\left(\mathbf{r_{ij}}\right)=\frac{\partial\mathcal{U}\left(\mathbf{r_{ij}}\right)}{\partial r_{ij}}\times\frac{1}{r_{ij}}\left(\mathbf{r_{j}-\mathbf{r_{i}}}\right) 
\end{equation}
where $\mathbf{r_{i}}$ and $\mathbf{r_{j}}$ are the positions of particles $i$ and $j$, $r_{ij}$ the distance between the two particles and $\mathcal{U}$ the potential energy.

\begin{figure}[ht!]
\centering{}\includegraphics[scale=0.5]{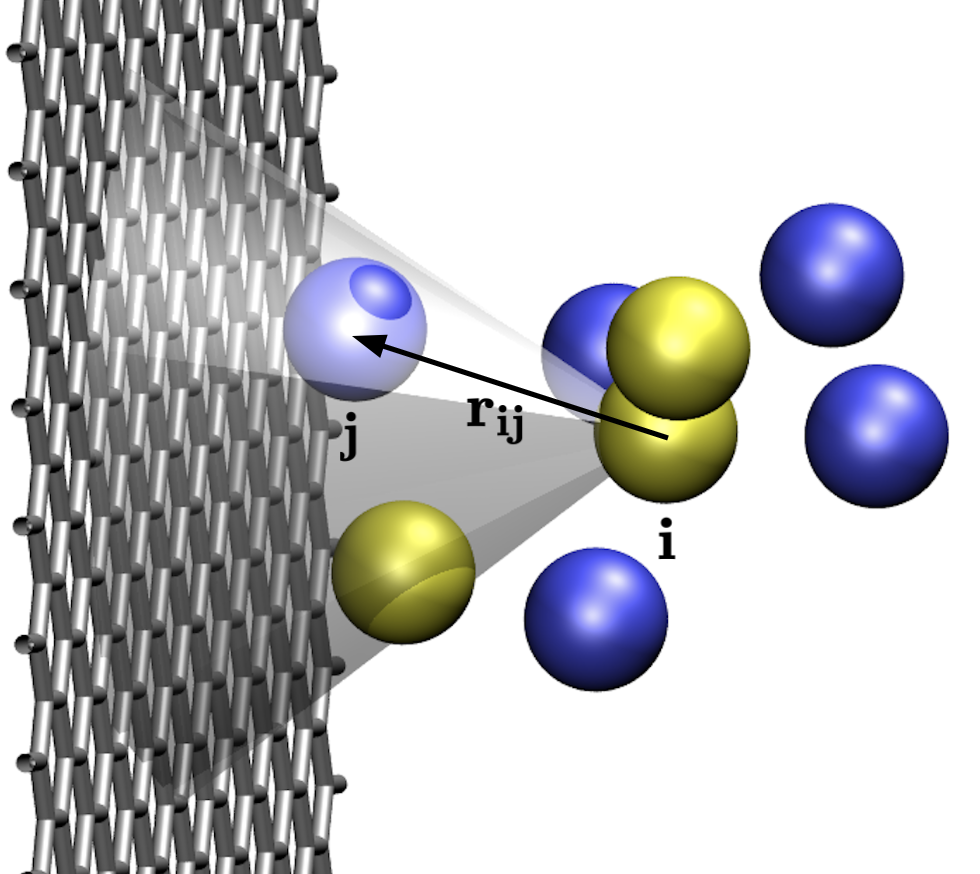}\caption{Scheme describing the definition of the neighborhood of an ion in this work. The two colors represent anions and cations. \label{fig:schema-voisinage}}
\end{figure}

In this analysis, ions are considered as single molecular entities (and not by their individual atoms) so the interactions between two ions are defined in terms of the interactions involving their respective centers of mass. Then, the position of a molecule is determined by the position of its centre of mass, while its charge is calculated as the sum of all its partial atomic charges. Since adsorption takes place along the $z$ direction and in order to account for an ion ``field of vision'', we define the neighborhood of an ion as an infinite cone oriented along $z$ (Figure~\ref{fig:schema-voisinage}). Only the centres of mass and carbon atoms in this neighborhood of the adsorbing ion are taken into account when calculating the contributions of the forces. The dimensions of the cone have been chosen so that at 9~{\AA} (corresponding to the first minimum of the EMIM-TFSI pair distribution function, see Figure~S5) the diameter of the cone is equal to 10~{\AA} (approximate size of an electrolyte ion).
Using the expression for the Coulomb and Lennard-Jones potential energies given in SI, the equations for the forces are derived. The Coulomb force exerted by the particle $j$ of charge $Q_{j}$ on the particle $i$ of charge $Q_{i}$ is given by
\begin{equation}
\mathbf{F_{ij}^{Coulomb}}=-\frac{1}{4\pi\varepsilon_{0}}\frac{Q_{i}Q_{j}}{r_{ij}^{3}}\left(\mathbf{r_{j}-\mathbf{r_{i}}}\right)
\end{equation}
where $\varepsilon_{0}$ denotes the permittivity of vacuum. The Coulomb force between an ion in the electrolyte and an atom in the electrode is given by
\begin{equation}
\mathbf{F_{ij}^{Coulomb}}=-\frac{1}{4\pi\varepsilon_{0}}\frac{Q_{i}Q_{j}}{r_{ij}^{3}}\left(\mathrm{erf\left(\textrm{\ensuremath{\eta\,r_{ij}}}\right)}-\frac{2}{\sqrt{\pi}}\eta\,r_{ij}{\rm e}^{-\left(\eta\,r_{ij}\right)^{2}}\right)\left(\mathbf{r_{j}-\mathbf{r_{i}}}\right)
\end{equation}
where $\mathrm{erf}$ is defined as the primitive of a Gaussian function. The force derived from the Lennard-Jones potential, describing the van der Waals interactions, is given by:
\begin{equation}
\mathbf{F_{ij}^{vdW}}=24\varepsilon_{ij}\left(\frac{\sigma_{ij}^{12}}{r_{ij}^{14}}-2\frac{\sigma_{ij}^{6}}{r_{ij}^{8}}\right)\left(\mathbf{r_{j}-\mathbf{r_{i}}}\right)
\end{equation}

To work in the centre of mass representation and in order to compute $\mathbf{F_{ij}^{vdW}}$, it is necessary to define a new set of parameters $\varepsilon_{ii}$ and $\sigma_{ii}$ corresponding to anions and cations. Following the method described in SI, the obtained values are $\sigma_{\mathrm{EMIM}}=4.02\,\textrm{{\AA}}$, $\sigma_{\mathrm{TFSI}}=5.45\,\textrm{{\AA}}$, $\varepsilon_{\mathrm{EMIM}}=0.71~\mathrm{kJ~mol^{-1}}$ and $\varepsilon_{\mathrm{TFSI}}=1.05~\mathrm{kJ~mol^{-1}}$.

The $z$ component of the total force exerted on the centre of mass of the ions adsorbing during the charging of the supercapacitor is then given by 
\begin{equation}
\mathbf{F_{ij}^{tot}.e_{z}}\left(t'\right)=\underset{j}{\sum}\left(\mathbf{F_{ij}^{Coulomb}}\left(t'\right)+\mathbf{F_{ij}^{vdW}}(t')\right)\mathbf{.e_{z}}
\end{equation}
where $j$ is the $j-th$ particle contained in the cone (Figure \ref{fig:schema-voisinage}) and $\mathbf{e_{z}}$ is a unit vector along~$z$.
\begin{figure}[ht!]
\centering{}\includegraphics[scale=0.5]{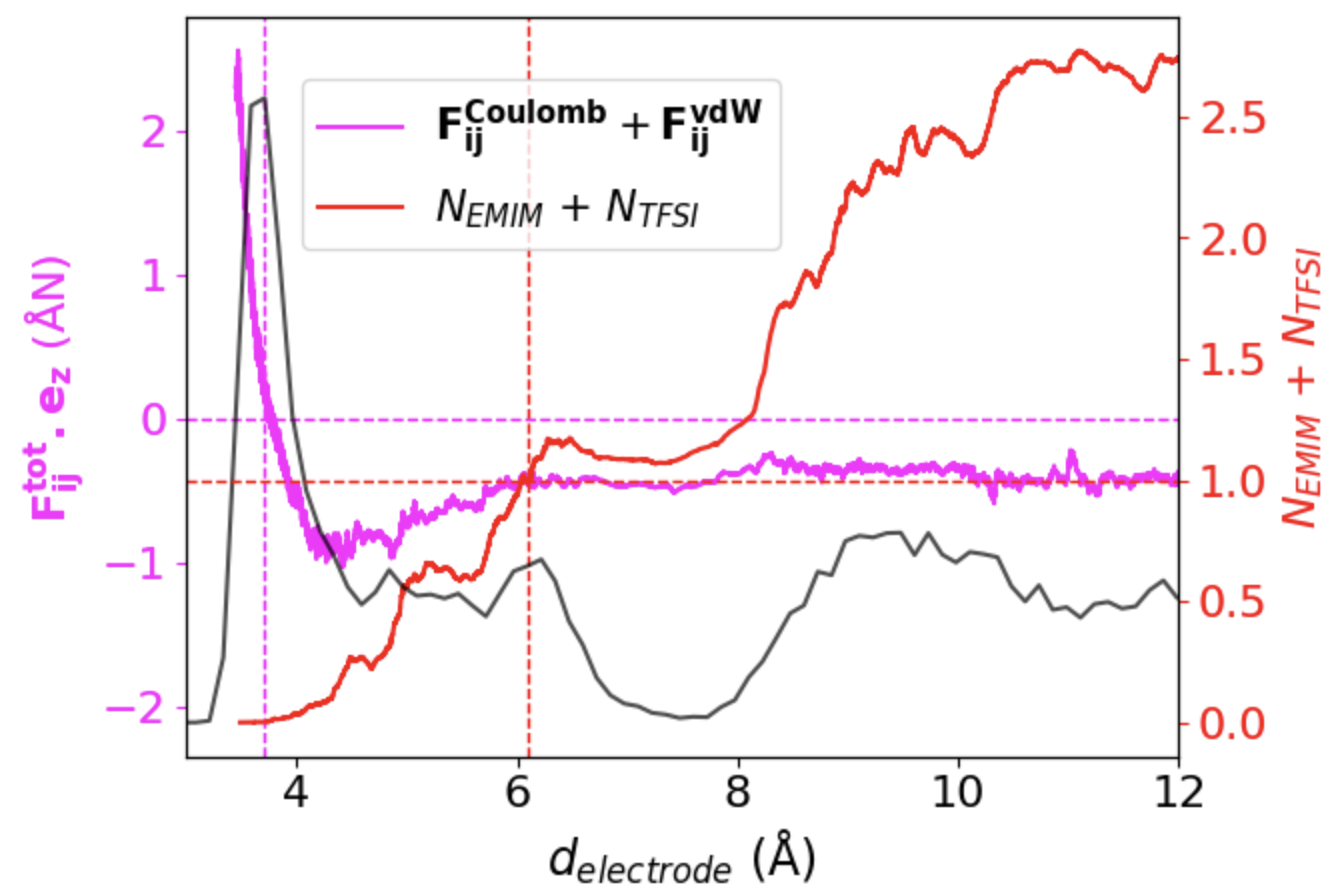}\caption{$z$-component of the mean total force felt by the adsorbing EMIM (in magenta) and number of neighboring EMIM and TFSI ions (in red) as a function of the distance from the electrode for $\Delta\Psi=$1V. The horizontal dashed lines highlight i) where the total force is zero and ii) the frontier where total number of ions in the cone is exactly 1. The solid black line is the cation density profile (arbitrary units) as a guide for the eyes to know which layers correspond to given distance $d_{electrode}$.
\label{fig:F_tot_nbr_coordo_vs_dist_elec}}
\end{figure}
Figure~\ref{fig:F_tot_nbr_coordo_vs_dist_elec} shows the mean total force felt by the EMIM during adsorption for $\Delta\Psi=1\mathrm{\,V}$ as a function of the distance from the electrode. For the negative electrode, the EMIM cations adsorb in the direction opposite to the axis $z$. So, if the force is directed toward the electrode, then $\mathbf{F_{ij}^{tot}.e_{z}}<0$ and then the force is attractive. Conversely, if $\mathbf{F_{ij}^{tot}.e_{z}}>0$ the force is repulsive. 

Different regimes can be identified in Figure~\ref{fig:F_tot_nbr_coordo_vs_dist_elec}. The force is i) attractive and constant for $d_{electrode}$ between 6.1 and 12~\r{A}, ii) increasing linearly for $d_{electrode}$ between 4.1 and 6.1~\r{A}, iii) decreasing and becoming strongly repulsive for $d_{electrode}$ smaller than 4.1~\r{A}. The distance from the electrode at which a maximum ion density is observed corresponds to the distance at which the total force cancels out. At this distance, the force is neither repulsive nor attrac\-tive which explains this maximum of density. In addition, it should be noted that the regime change observed at $d_{electrode}\sim4.1\,\textrm{{\AA}}$, coincides with the time at which the ions have, on average, less than one molecule in their field of vision. This shows that a vacant space must be available to facilitate the adsorption process. Additional plots showing the results of the same analysis with TFSI and the statistical uncertainty from the average curves are available in Figures~S6, S7 and~S8. 

The kinetic energy of an ion is $E_{K}=\frac{1}{2}m\left(v_{x}^{2}+v_{y}^{2}+v_{z}^{2}\right)$ where $v_{\alpha}$ correspond to the different components of the ion velocity. For $d_{electrode}\gtrsim4.1\,\textrm{{\AA}}$ and $d_{electrode}\lesssim8\,\textrm{{\AA}}$, $v_{x}^{2}$ and $v_{y}^{2}$ are close to zero, so the kinetic energy behaves like $v_z^2$. Figure S9 shows that the component of the total force along the $x$-axis and the $y$-axis is zero, which is an additional argument for looking only along the $z$-axis. 
\begin{figure}[ht!]
\centering{}\includegraphics[scale=0.5]{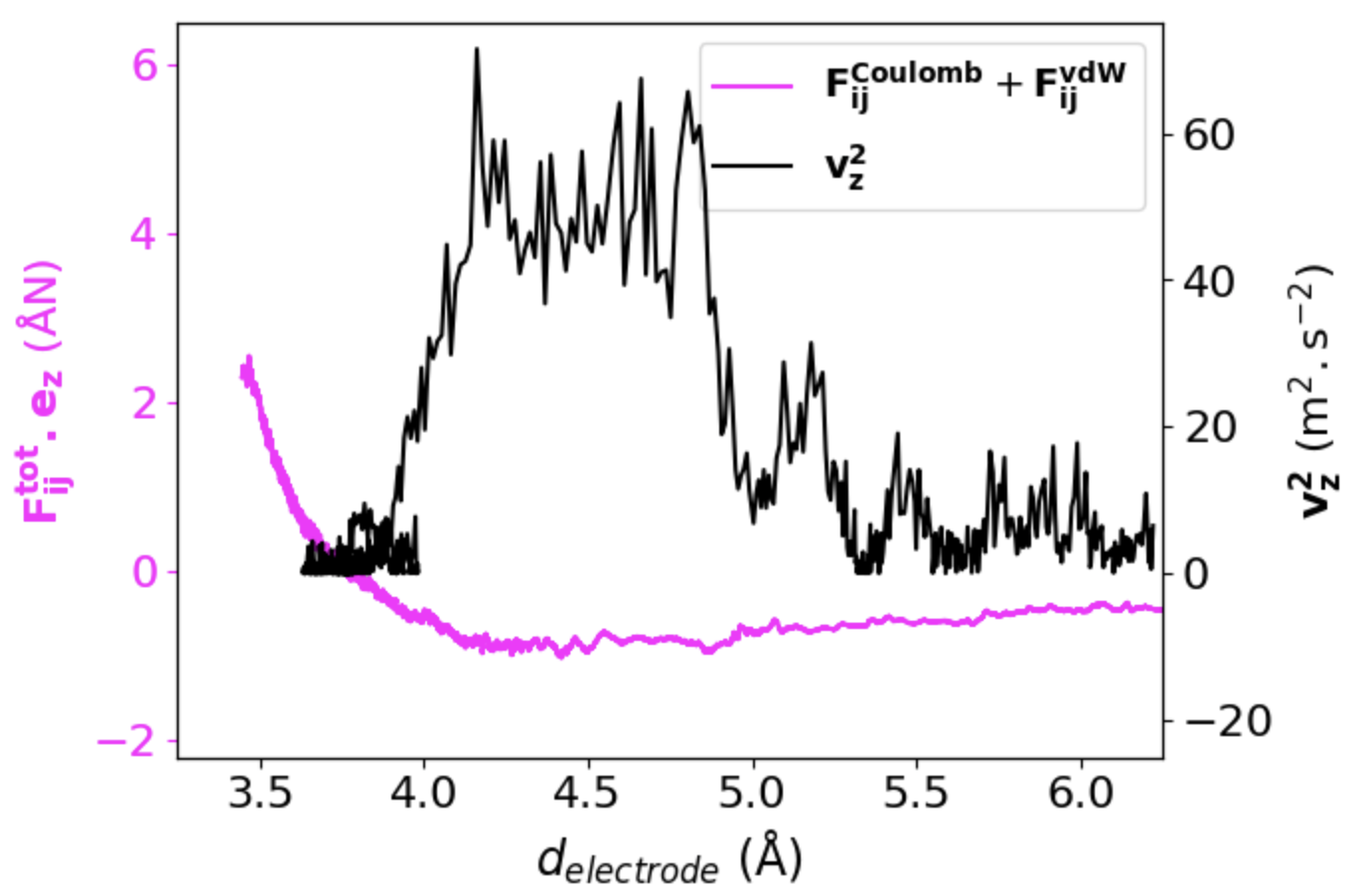}\caption{$z$-component of the mean total force felt by the adsorbing EMIM (in magenta) and a quantity proportional to the kinetic energy ($\frac{1}{2}\left\langle v_{z}\right\rangle ^{2}$, in black) as a function of the distance from the electrode for $\Delta\Psi=$1V. The black dashed line correspond to the distance from the electrode where $\mathbf{F_{ij}^{tot}.e_{z}}=0$. \label{fig:kinetic-energy}}
\end{figure}
For this reason, on Figure~\ref{fig:kinetic-energy} we compare the total force felt by the ion with $v_z^2$ as a function of the ion-electrode distance. Figure~\ref{fig:kinetic-energy} shows that the kinetic energy reaches its maximum at the distance for which the force is the most attractive, i.e. for $d_{electrode}\approx z_{ads}.$ In the neighborhood of the distance for which $\mathbf{F_{ij}^{tot}.e_{z}}=0$, the force does not modify the kinetic energy, the work of the force is zero: the ions diffuse in the plane of the electrode.
\begin{figure}[ht!]
\begin{centering}
\includegraphics[scale=0.5]{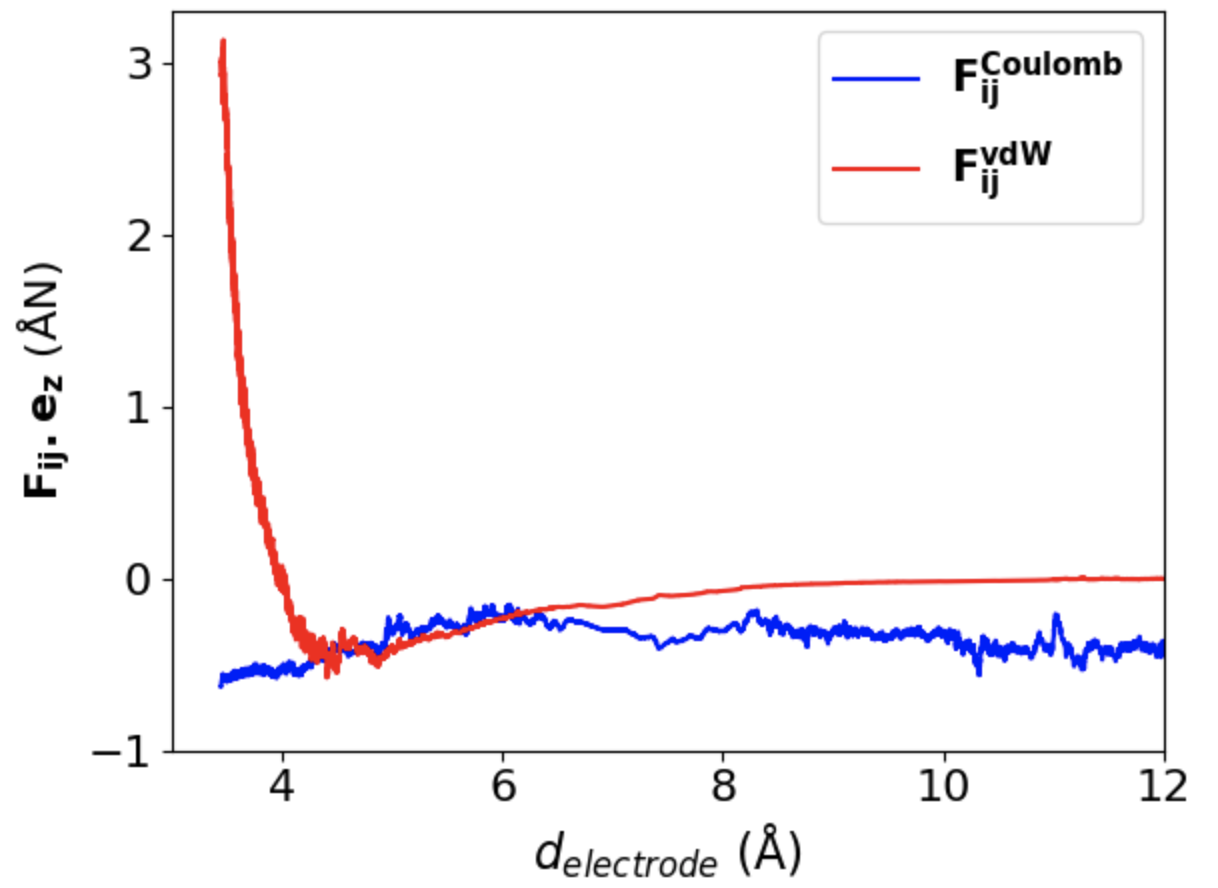}\caption{$z$-component of the van der Waals (in blue) and Coulomb forces (in red) as a function of the distance
from the electrode for $\Delta\Psi=$1V.\label{fig:F_LJ_Vs_Coul}}
\par\end{centering}
\end{figure}

Figure \ref{fig:F_LJ_Vs_Coul} shows the breakdown of the contributions of the van der Waals and Coulomb forces to the adsorption process. The Coulomb force is attractive over the whole set of distances, which is due to the presence of an opposite charge on the surface of the electrode. At long distances this attractive Coulomb force is the major contribution to the total force and it drives the adsorption of the ion. At short distances a larger contribution from the van der Waals force is observed, which is due to the Pauli repulsion between the electronic clouds, and prevent the ion to approach too close to the electrode. Finally, it is worth noting that in the intermediate region, for $d_{electrode}$ between 4.1 and 6.1~\r{A} the two contributions are positive and almost equal.

\begin{figure}[ht!]
\begin{center}
\includegraphics[scale=0.36]{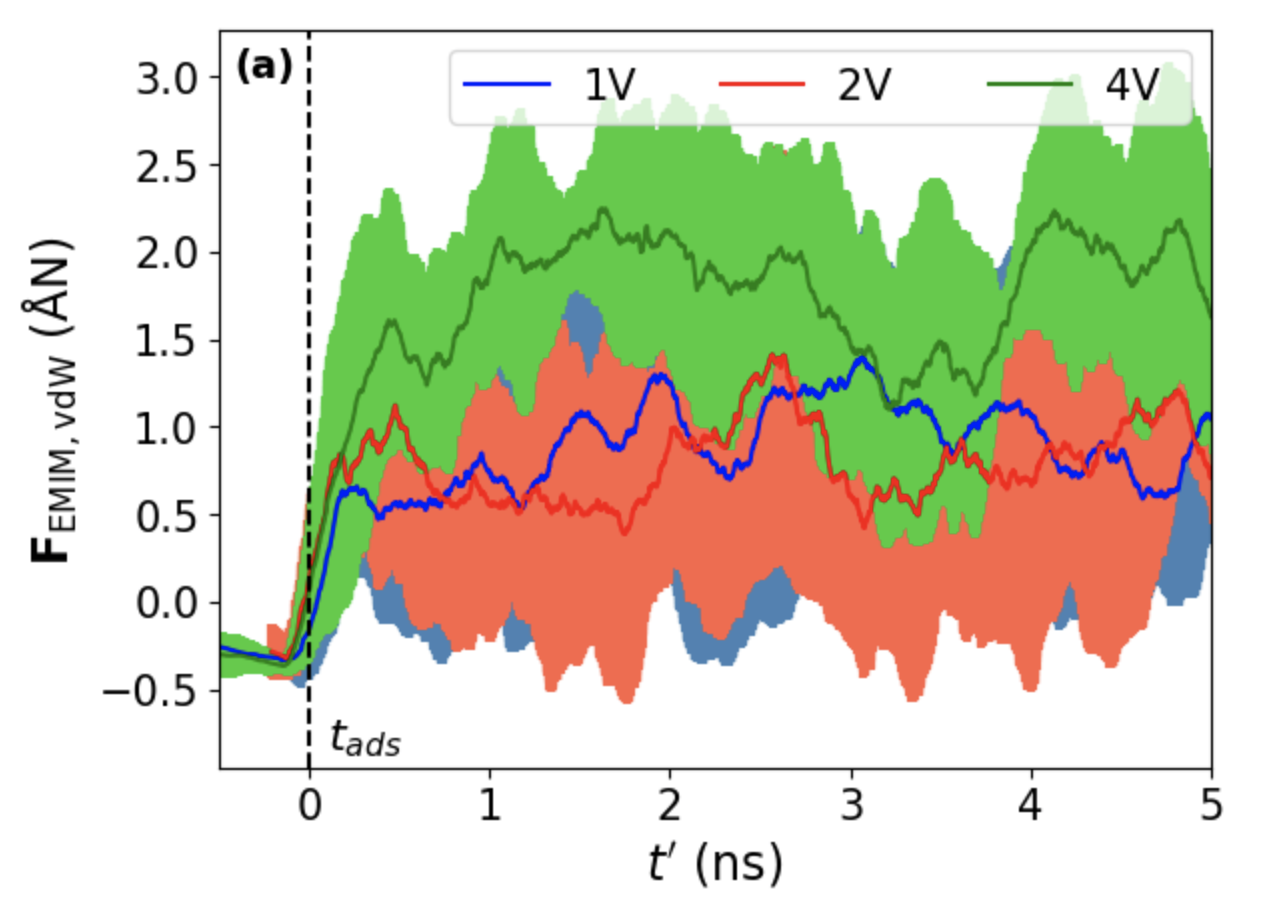}
\includegraphics[scale=0.36]{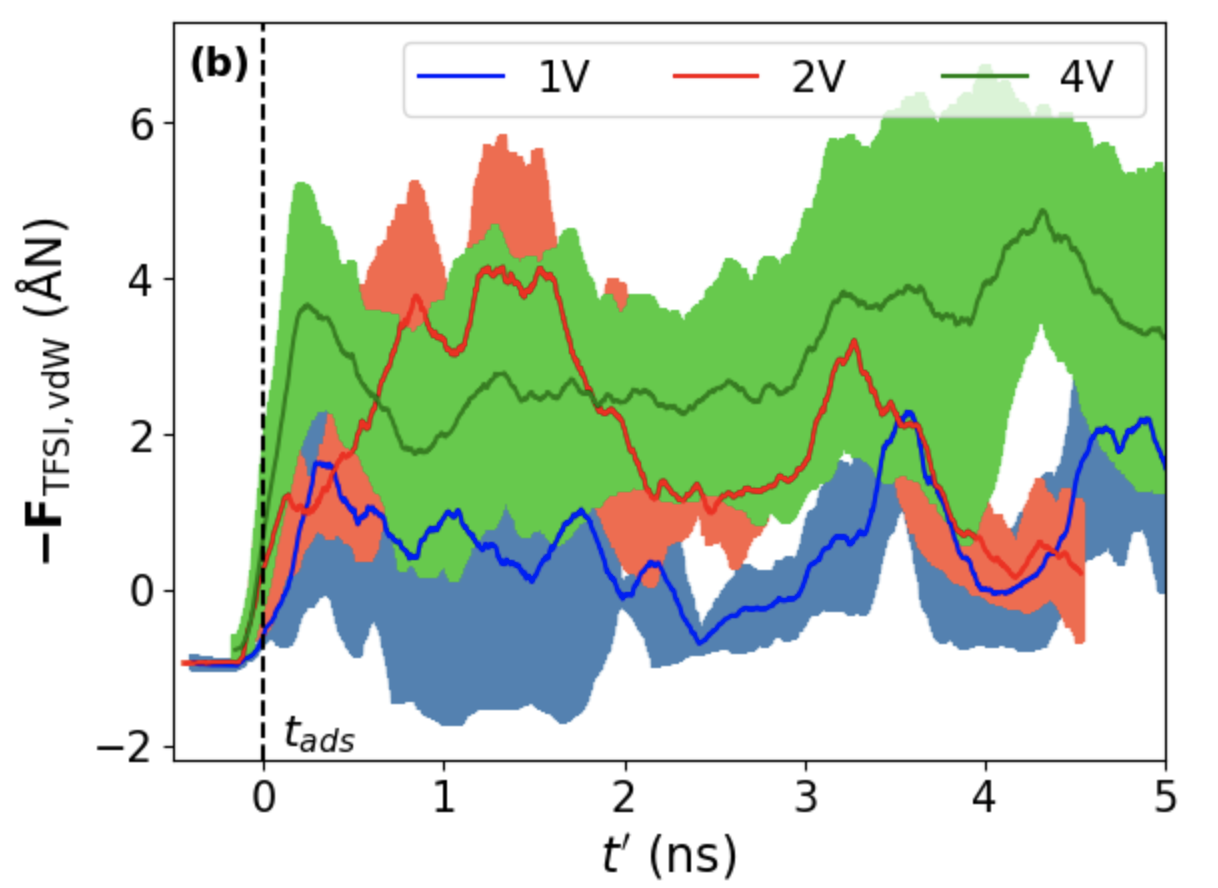}
\caption{Average van der Waals force felt by EMIM (a) and TFSI (b) ions during adsorption over time for $\Delta\Psi=1,\,2,\,4~\mathrm{V}$. Note that the y-axis scales are different.\label{fig:Average_F_LJ_diff_pot}}
\end{center}
\end{figure}

Figure~\ref{fig:Average_F_LJ_diff_pot} shows the van der Waals force felt by the adsorbing ions as a function of time for $\Delta\Psi=1,\,2,\,4~\mathrm{V}$. It can be seen that before being adsorbed, the forces experienced by the ions do not exhibit any variation with respect to the applied potential and remain consistently attractive. This is expected since this force, by definition, is not contingent upon charge. 

Once the ions are adsorbed, the force is essentially repulsive but the standard deviations of the curves are such that it is difficult to draw conclusions. Indeed, as seen in Figure~\ref{fig:F_LJ_Vs_Coul}, once the ions are adsorbed, for small variations in distance there is a large variation in force. Nevertheless, if one only considers the mean value, it seems that the magnitude of the repulsive forces increases with the applied potential difference. This is due to the ion-electrode distance decreasing with an increasing applied potential, which consequently increases the force. Furthermore, the fact that $\varepsilon_{\mathrm{EMIM}}$ is smaller than $\varepsilon_{\mathrm{TFSI}}$ can explain why $\left|\mathbf{F_{EMIM}^{LJ}.e_{z}}\right|$ is smaller than $\left|\mathbf{F_{TFSI}^{LJ}.e_{z}}\right|$.

\begin{figure}[ht!]
{\raggedright{}\includegraphics[scale=0.4]{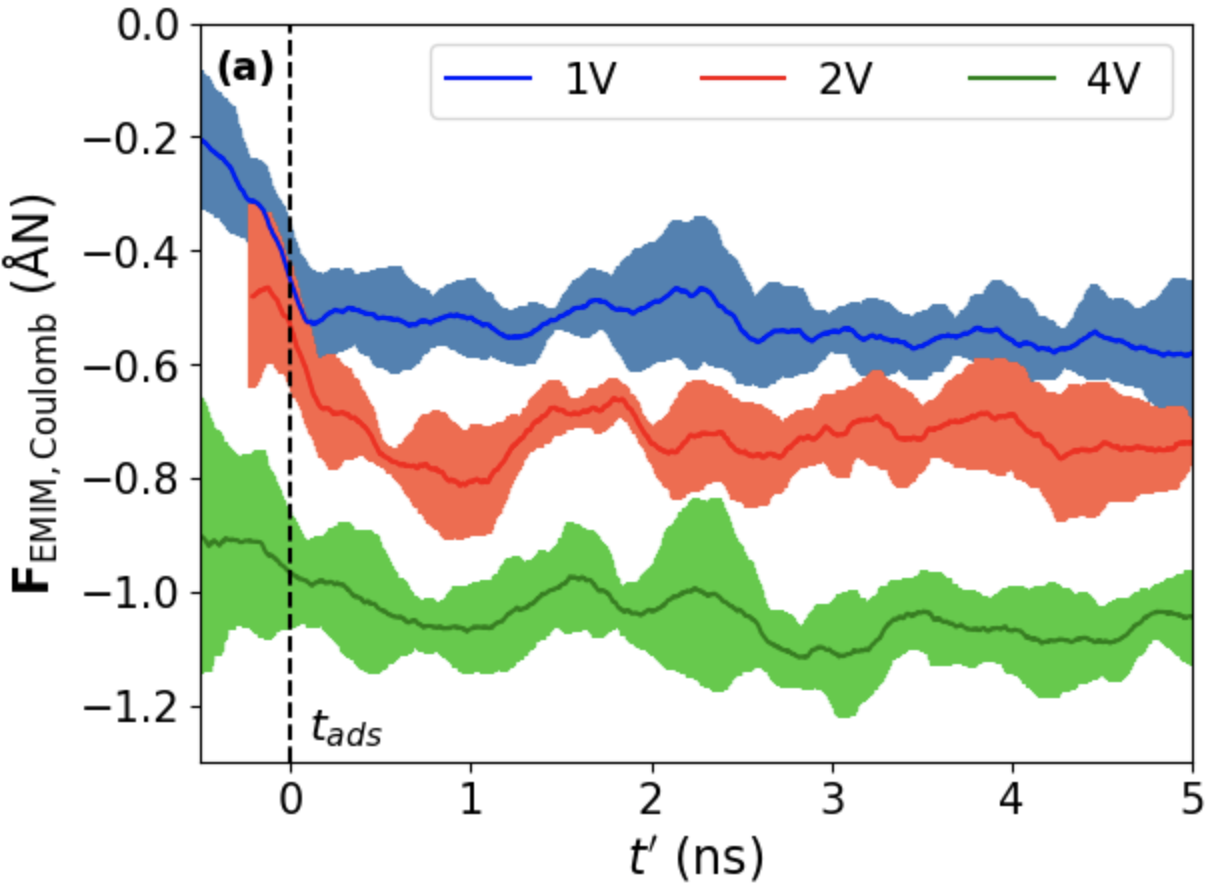}}{\raggedleft{}\includegraphics[scale=0.4]{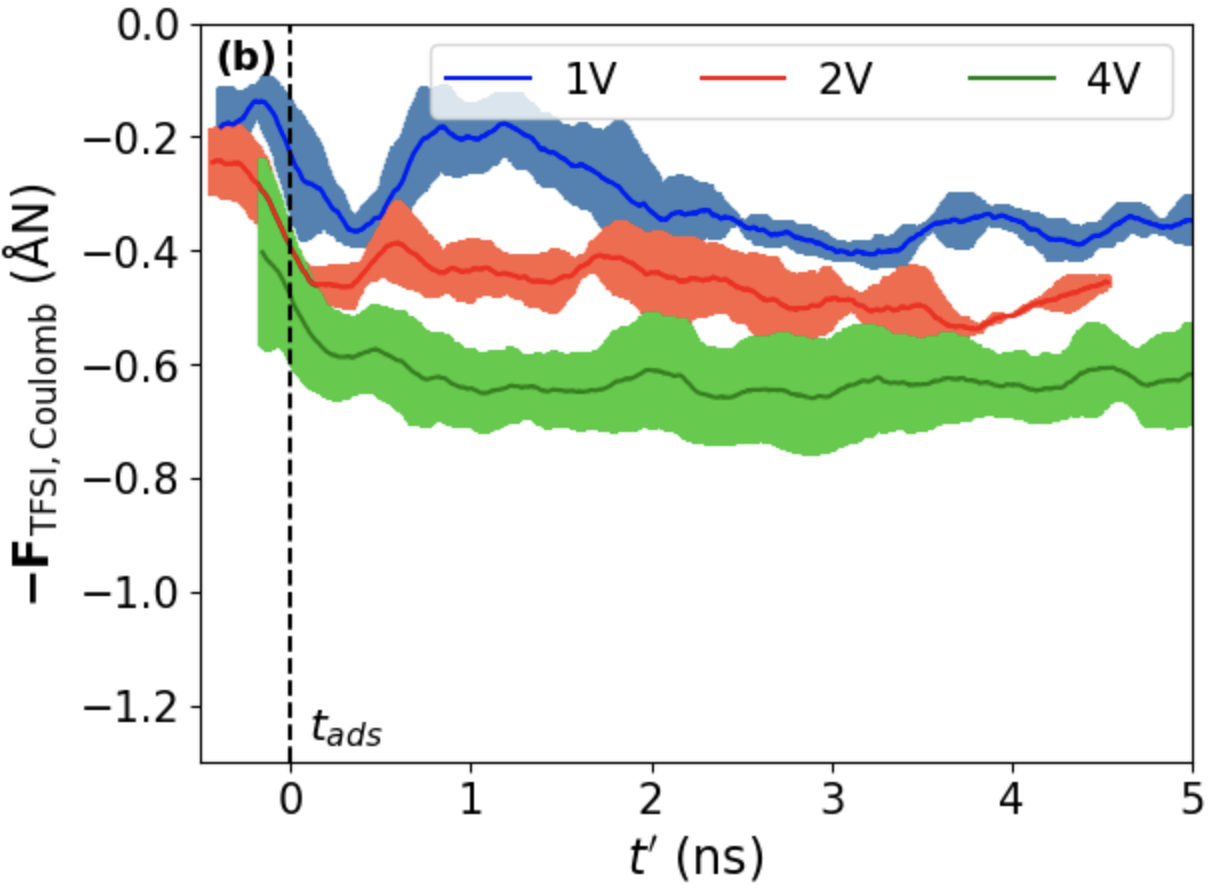}}\caption{Average Coulomb force felt by EMIM (a) and TFSI (b) ions during adsorption over time for potentials $\Delta\Psi=1,2~\mathrm{and}~4~\mathrm{V}$.\label{fig:Average-Coulomb-force_diff_pot}}
\end{figure}

Figure~\ref{fig:Average-Coulomb-force_diff_pot} shows the Coulomb force felt by the adsorbing ions for $\Delta\Psi=1,\,2,\,4~\mathrm{V}$. For all applied potentials the forces are attractive and increase at the time of adsorption. Furthermore, the attractiveness of the force increases with the potential. This observation is consistent with the idea that when the potential increases, the charge carried by the electrodes also increases. Consequently, this increased charge amplifies the Coulomb force experienced by the ions, intensifying the attraction between the ions and the electrodes.
The force felt by the cations is greater compared to the force felt by the anions. This can be explained by the fact that after adsorption, the centres of mass of the cations are on average closer to the electrode than for the anions (see Figure~S10). It is worth noting that the quantitative results presented here may depend on the nature of the neighborhood considered. 

\begin{figure}[ht!]
\begin{center}
\includegraphics[scale=0.36]{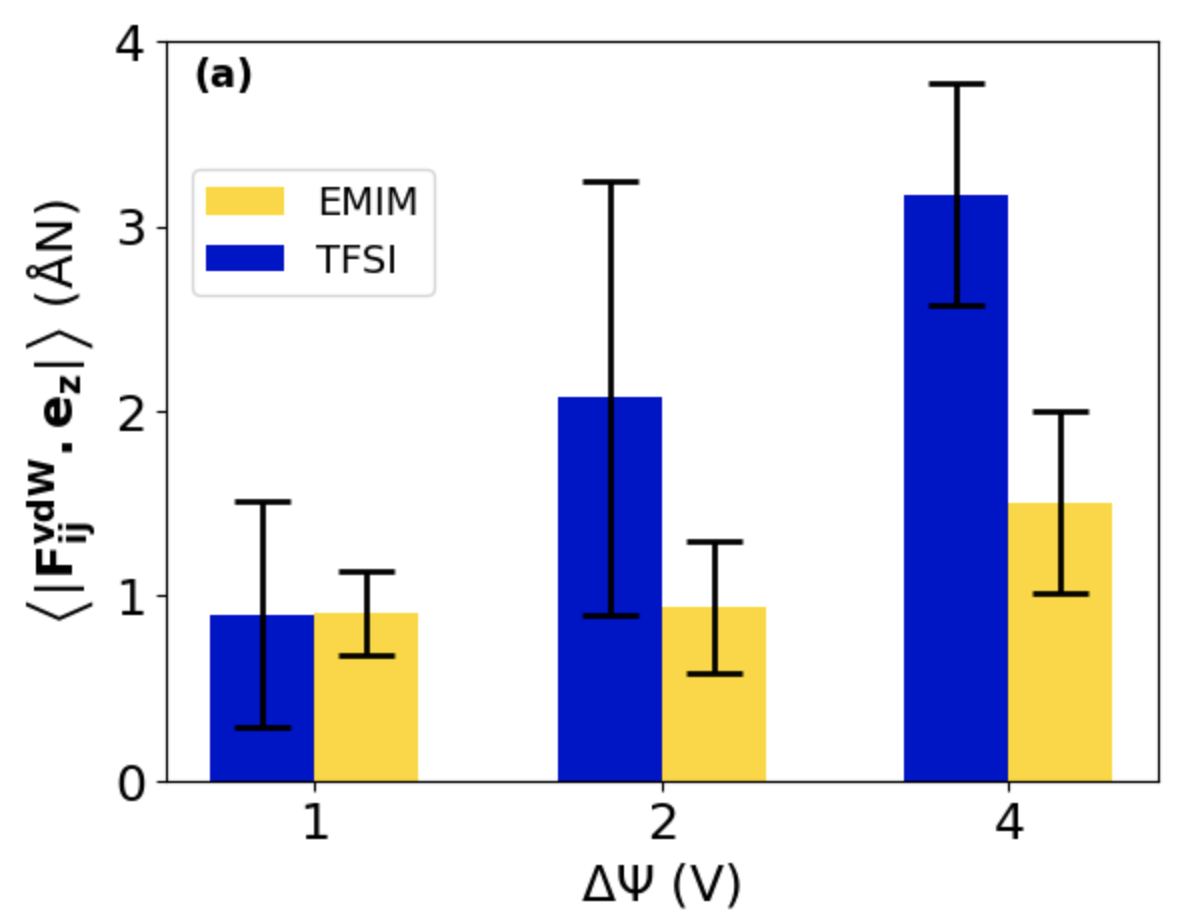}
\includegraphics[scale=0.36]{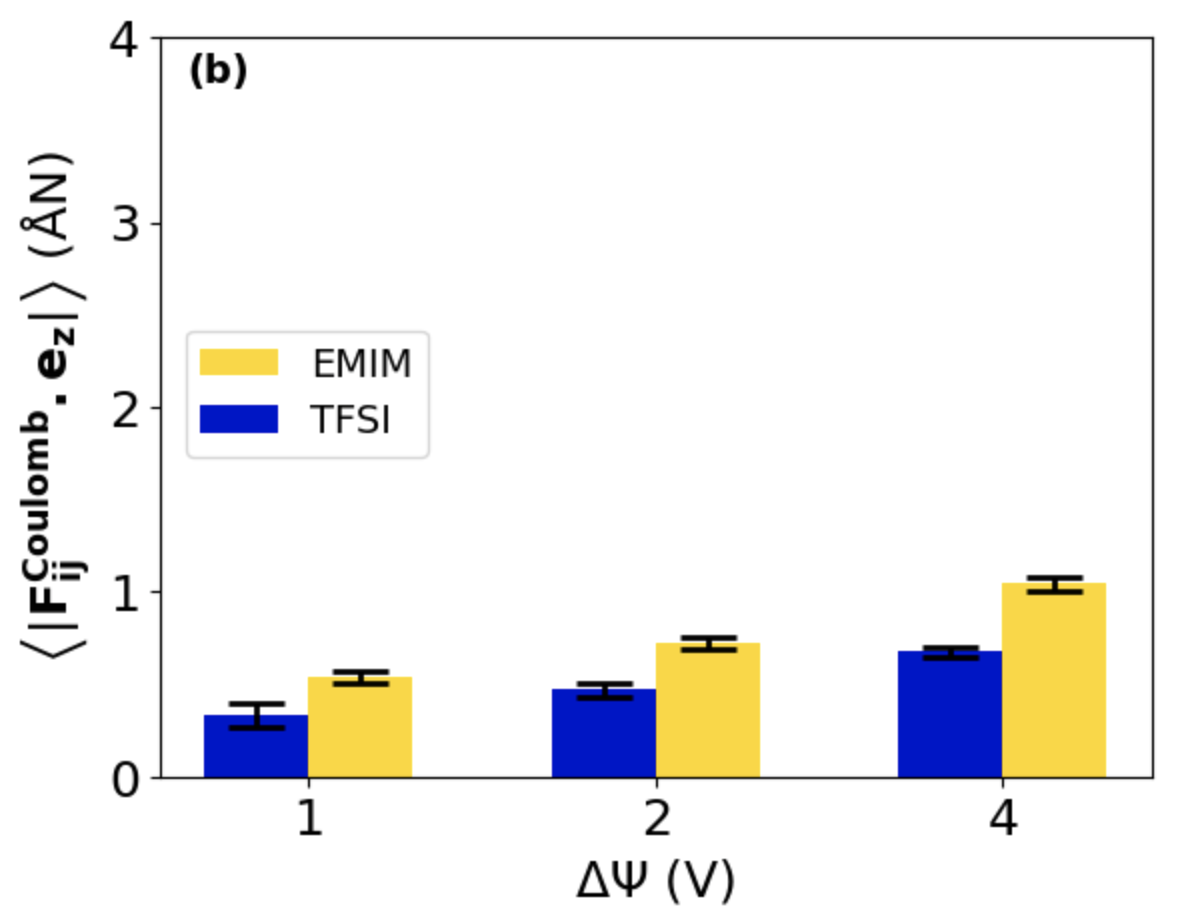}
\caption{Average of the absolute van der Waals force felt by EMIM and TFSI (a) and average of the absolute Coulombic force felt by EMIM and TFSI (b) after adsorption for $\Delta\Psi=1,\,2,\,4~\mathrm{V}$. Averages were done after ions are adsorbed, for times $t'\geq1$.\label{fig:Average_F_LJ_and_coulomb_diff_pot}.}
\end{center}
\end{figure}

The comparison of the average absolute values of forces shown in Figure~\ref{fig:Average_F_LJ_and_coulomb_diff_pot} reveals that, at equilibrium, the van der Waals forces exhibit greater magnitude than the Coulomb forces. This is consistent with the recent findings reported by de Araujo Chagas \textit{et al} with graphene/graphyne electrodes and {[}EMIM{]}{[}BF4{]} electrolyte\cite{de_araujo_chagas_comparing_2023}.
This figure also shows clearly the dependance of the forces on the applied potential.

\section{Conclusion}

The literature concerning the description, through molecular dynamics simulations, of the EDLs formed by ionic liquids in supercapacitors focused mostly on equilibrium results so far. Most of the studies present the ion densities and ion orientations for different electrodes and electrolytes in order to compare the capacitive properties of the simulated devices. However, an understanding of the adsorption and charging dynamics of supercapacitors is needed to improve the performance of such systems. So we have developed methods for generalising and dynamically describing the ion adsorption process. 

In this work, we focused on a typical supercapacitor, made of two graphite electrodes separated by pure ionic liquid {[}EMIM{]}{[}TFSI{]} for potential differences of $\Delta\Psi=1,\,2,\,4~\mathrm{V}$. Using molecular dynamics, we studied the dynamics of EDL creation by describing the orientation of adsorbing ions in relation to their distances from the electrode, and by analyzing the forces acting on them.

Firstly, we have seen that EMIM and TFSI ions reorientate as they adsorb, regardless of the potential applied. The closer the ions are to the electrode, the more they tend to be parallel to it. We have seen that this tendency varies linearly with the distance from the electrode, {\it i.e.} reorientation occurs simultaneously to the adsorption.

Secondly, we described and detailed the contributions of the van der Waals and Coulomb forces acting on the ions during adsorption. We have seen that Coulomb forces dominate at long range while van der Waals forces dominate at short range. In addition, we saw that there was an almost equal contribution from the two forces at an intermediate distance. Interestingly, at short range, both contributions can depend on the potential applied, as a consequence of the reduced ion-electrode distance.

In the future, the method developed here could be applied to more complex materials, such as MoS$_2$,~\cite{bi2022a} MXenes,~\cite{Xu20} etc, for which the presence of a variety of elements may change the balance between the main interactions, leading to different adsorption processes. The method could further be used to explore the case of nanoporous materials, but it would be necessary to account for their complex geometries by adapting dedicated tools~\cite{dick2023a} in particular when detailing the forces experienced by the ions during adsorption.

\begin{acknowledgement} 

This project has received funding from the European Research Council (ERC) under the European Union's Horizon 2020 research and innovation program (grant agreement no. 714581). This work was granted access to the HPC resources of TGCC under the allocations AD010911061R1 and A0110911061 made by GENCI. The authors acknowledge Camille Bacon for initial simulation files.

\end{acknowledgement}

\begin{suppinfo}

Supplementary Information available: Additional details on the methods for defining adsorption events and finding a common time scale, additional details on the definitions of energy and coarse-grained Lennard-Jones parameters, evolution of the total charge as a function of time, raw data for the orientation of ions and ion-electrode distance,
pair distribution functions between ions, forces for TFSI ions and additional data on forces, average ion-electrode distances.

\end{suppinfo}

\section*{Data availability}

The data corresponding to the plots reported in this paper, as well as example input files for the MD simulations, are available in the Zenodo repository with the identifier 10.5281/zenodo.10695593.

\providecommand{\latin}[1]{#1}
\makeatletter
\providecommand{\doi}
  {\begingroup\let\do\@makeother\dospecials
  \catcode`\{=1 \catcode`\}=2 \doi@aux}
\providecommand{\doi@aux}[1]{\endgroup\texttt{#1}}
\makeatother
\providecommand*\mcitethebibliography{\thebibliography}
\csname @ifundefined\endcsname{endmcitethebibliography}
  {\let\endmcitethebibliography\endthebibliography}{}

\end{document}